\documentclass[ag]{copernicus}

\usepackage{color}

\begin{document}

\title{Collisionless magnetic reconnection: Flux quanta, field lines, `composite electrons' -- Is the quantum-Hall effect involved in its micro-scale physics?}

\author[1,2]{R. A. Treumann\thanks{Visiting the International Space Science Institute, Bern, Switzerland}}
\author[3]{R. Nakamura}
\author[3]{W. Baumjohann}

\affil[1]{Department of Geophysics and Environmental Sciences, Munich University, Munich, Germany}
\affil[2]{Department of Physics and Astronomy, Dartmouth College, Hanover NH 03755, USA}
%\affil[3]{International Space Science Institute, Bern, Switzerland}
\affil[3]{Space Research Institute, Austrian Academy of Sciences, Graz, Austria}

\runningtitle{Micro-scale physics of collisionless reconnection}

\runningauthor{R. A. Treumann, R. Nakamura, and W. Baumjohann}

\correspondence{R. A.Treumann\\ (rudolf.treumann@geophysik.uni-muenchen.de)}

\received{ }
%\pubdiscuss{ } %% only important for two-stage journals
\revised{ }
\accepted{ }
\published{ }

%% These dates will be inserted by the Publication Production Office during the typesetting process.

\firstpage{1}

\maketitle

\begin{abstract}
Microscopically, collisionless reconnection in thin current sheets is argued to involve `composite electrons' in the ion inertial (Hall current) domain, a tiny fraction of electrons only. These `composite electrons' are confined to lower Landau levels $\epsilon_L\ll T_e$ (energy much less than temperature). They demagnetise by absorbing magnetic flux quanta $\Phi_0=h/e$, decouple from the magnetic field, transport the attached magnetic flux into the non-magnetic centre of the current layer, where they release the flux in the form of micro-scale magnetic vortices, becoming ordinary electrons. The newly born micro-scale magnetic vortices reconnect in their strictly anti-parallel sections when contacting other vortices, ultimately producing the meso-scale reconnection structure. We clarify the notions of magnetic field lines and  field line radius, estimate the power released when two oppositely directed flux quanta annihilate, and calculate the number density and Landau-level filling-factor of `composite electrons' in the Hall domain. As side product we find that the magnetic diffusion coefficient in plasma also appears in quanta $D_0^m=e\Phi_0/m_e=h/m_e$, yielding that the bulk perpendicular plasma resistivity is quantised, with quantum (lowest limit) $\eta_{\,0\perp}=\mu_0 e\Phi_0/m_e=\mu_0h/m_e\sim 10^{-9}$ Ohm m. 

\end{abstract}

\introduction
It is the intuitive simple geometric picture of annihilating antiparallel magnetic field lines when  approaching each other  \citep[see, e.g.,][]{sweet1957,parker1958,dungey1961} that led to the idea of magnetic reconnection as the plasma process that converts stored magnetic energy into kinetic energy. The approach of the fields in these models is assumed to be provided by the encounter of two magnetised plasma streams which have been treated as electrically conducting fluids. Observations in space  have  confirmed the presence of reconnection under completely collisionless conditions \citep[see, e.g.,][]{fujimoto1997,nagai2001,oieroset2001} when the classical fluid theoretical approaches break down. The first observational confirmation of reconnection in space under collisionless conditions dates back to \citet{paschmann1979} who detected the so-called `reconnection jets' which were ejected about symmetrically from the local reconnection site at Earth's magnetopause, although the conditions were neither identified nor noted as being collisionless at the time. The reconnection site could not be resolved (and has never been resolved) and the very process of reconnection itself that was going on could not be identified, neither in these observations nor in any spacecraft observations that followed until today. It was attributed to some kind of unidentified diffusion process that would be able to let the magnetic field `diffuse' across the plasma in order to merge, annihilate and subsequently reconnect. 

Previous theoretical models were, with only a few exceptions \citep[][]{galeev1975,sonnerup1979} based on diffusive fluid approaches invoking mass and momentum conservation of fluid elements \citep[cf., e.g.,][for the two kinds of early canonical treatment]{parker1958,petschek1964} across the reconnection site. The magnetic field in these fluid models was subject to the induction equation that included some form  of  finite electrical conductivity $0<\sigma\neq\infty$, either distributed or localised,  resulting from the fluid-momentum conservation equation of the particles (essentially only the electrons, because of their high mobility). In the initial approaches to reconnection physics, only resistive (or anomalously resistive) interaction was assumed of enabling reconnection but was later extended to include also other electron moments as pressure gradients, pressure tensors, ion viscosity, Hall terms, nonlinear momentum terms, and finally ponderomotive forces generated by nonlinear wave interactions \citep[see, e.g.,][for a discussion of some of these effects]{biskamp2000}. 

A seminal review of the early fluid models \citep{vasyliunas1975} provided the expected macroscopic flow and field geometries but missed the important meso- and micro-scale effects caused by the mass difference between ions and electrons, as was realised only somewhat later \citep{sonnerup1979}.  They cause a Hall current in the ion-inertial region $\lambda_e<|z|<\lambda_i$(with $\lambda_{e,i}=c/\omega_{e,i}$ and $\omega_{\,i,e}=e\sqrt{N/\epsilon_0m_{\,i,e}}$ the respective electron and ion inertial lengths and plasma frequencies). Ultimately, though delayed by two decades, its presence was observationally confirmed by \citet{fujimoto1997}, \citet{nagai2001}, and \citet{oieroset2001}. 

The region $|z|<\lambda_e$, the `electron inertial' domain in the centre of the current layer, is believed to be the very site of reconnection \citep[cf., e.g.,][]{scudder2008}. There the electrons are non-magnetised, and no Hall currents flow. Recently, referring to nonlinear (ponderomotive) interaction, we suggested that, in this region, the Weibel instability is capable of generating seed-{\sf X} points \citep{treumann2010} which may ignite reconnection on larger scales and also in the presence of a guide magnetic field along the sheet current. 
%To be more precise, the effect of the ponderomotive force has recently been noted \citep{treumann2001} but never included into any model. Its effect still remains to be clarified. As long as the ponderomotive force derives from an equivalent scalar potential, as is the case in stationary and homogeneous electrostatic Langmuir and ion acoustic plasma turbulence \citep[cf. any textbook on nonlinear kinetic plasma theory, e.g.,][]{treumann1997} it does not contribute to the induction equation because its curl vanishes identically, an effect that corresponds to a generalised Coulomb gauge. 
Ponderomotive interactions naturally structure the plasma locally. In this way they produce plasma gradients and pressure variations, graininess, and affect the particle distributions, causing structure in phase space like electron and ion holes \citep[cf., e.g.,][]{newman2001}. In two-dimensional numerical PIC simulations of reconnection, and particular in the presence of guide fields, the observation of similar structures has been reported to occur \citep{drake2003}. They are important in the Weibel scenario  \citep[][]{treumann2010}  providing conditions under which seed-{\sf X} points are formed and reconnection may start.  

In collisionless numerical simulations, reconnection is artificially ignited, either by {\it ad hoc} imposing a seed {\sf X}-point on the current sheet \citep[cf., e.g.,][]{zeiler2000,drake2003} or by locally injecting an artificial resistivity. The ongoing search for the mechanism of spontaneous onset of collisionless reconnection points to the `missing micro-scale physics' in thin current sheets. We argued somewhere \citep{treumann2010} that numerical simulations of reconnection imposing seed-{\sf X} points correctly describe the `meso-scale physics' by skipping the brief Weibel phase and replacing it with its final state of nonlinear saturation. %On the other hand, any `anomalous' models of reconnection (resistive, viscous etc.) are put into question.

Nevertheless, meso-scale physics does not answer the question yet of why and in which way reconnection proceeds at all. By which process can a magnetic field pass from the Hall region into the centre of the current layer \citep[see the discussion in][]{baumjohann2010}?
What happens at the reconnection site after the Weibel instability has saturated and generated a seed-{\sf X} point? What is the mechanism that lets magnetic fields merge and reconnect? 

Answering these questions requires an inquiry into the micro-scale physics. 
As a first preliminary step  we will, in the following, try to clarify the microscopic conditions which should be taken into account. This requires first the clarification of what is meant microscopically by a magnetic field line and by merging, annihilating and reconnecting anti-parallel magnetic field lines.

\section{Magnetic field lines}
The geometric view of reconnection arises from the concept of magnetic field lines. Historically magnetic field lines have been introduced by the ingenious imagination of Michael Faraday in the 1830s in his attempt to visualise the direction of magnetic forces in air. In vacuum (air) or any dielectric medium, magnetic field lines have no substance. They simply relect the vector character of magnetic induction and demonstrate the direction of the magnetic field. This changes in magnetically active media like plasmas, in particular in collisionless plasmas where particles are bound to the magnetic field and their mutual dynamics cannot easily be decoupled. 

How could a field line be defined in this case? Classical attempts were based on the use of so-called field-line potentials $\alpha({\bf x}, t),\beta({\bf x}, t)$, playing the role of particular coordinates. However, they simply replace vector potentials when describing the `motion of field lines'. They do not add anything substantial to the physical concept of a field line nor to physical understanding. In some sense they are a different formulation of the frozen-in concept or, on a more fundamental level, of the Lorentz transformation in ideal collisionless media. 

In order to arrive at a deeper understanding, one has to acknowledge that the definition of field lines as physical entities makes sense only in magnetically active matter, i.e. for charged particles which are sensitive to the presence of magnetic flux, for instance. 
\begin{figure}[t!]
\centerline{{\includegraphics[width=0.4\textwidth,clip=]{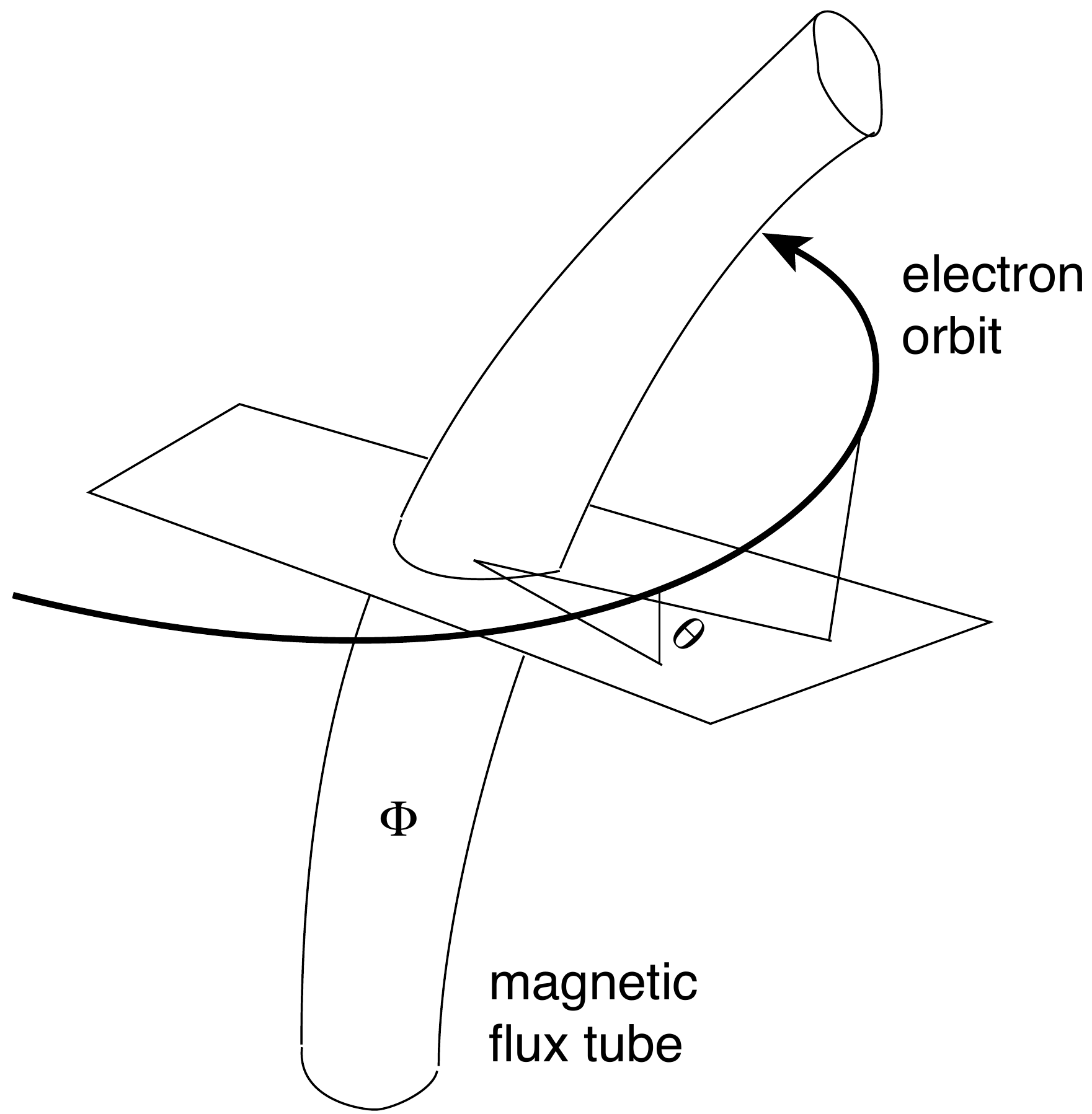}
%\centerline{{\includegraphics[width=0.4\textwidth,clip=]{weibel-fig2a.eps}
}}
\caption[ ]
{\footnotesize Geometry of the Aharonov-Bohm effect. The electron moves at some angle $\theta$ around the magnetic flux tube containing the magnetic flux $\Phi$. Since outside the flux tube the magnetic flux vanishes, the vector potential in this region reduces to a scalar potential which can be gauged away. However, quantum mechanically its effect is remarkable, forcing the flux to be quantised.}\label{fig-1}
\vspace{-0.3cm}
\end{figure}

\subsection{Field lines are flux quanta}
Under this proposition, a clear definition of a field line can indeed be derived from a quantum-mechanical treatment of magnetic fields in active matter. There is no need to explicitly solve the Schr\"odinger equation of  motion of an electron in a magnetic field as this was done long ago, in fact more than half a century ago, by \citet{landau1930} and \citet{aharonov1959}.\footnote{\citet{landau1930} treated the quantum-mechanical motion of an electron in a homogeneous magnetic field. \citet{aharonov1959} considered electrons moving around homogeneous flux tubes.  It is quite surprising that these attempts have not become to be known in space plasma physics, where magnetic field lines belong to the daily vocabulary and everyone dealing with reconnection, for instance, or dynamo theory and magnetic field models is using the concept of field line dynamics.} Gauging  the vector potential ${\bf A}$ by  adding the gradient of a scalar potential $-\nabla\phi$ in this case translates into a phase factor in the state eigenfunction $\psi({\bf x},t)$, the solution of the Schr\"odinger equation, 
\begin{equation}
\psi({\bf x},t,\phi) \sim \exp\left(-i\frac{e\phi}{\hbar}\right)\psi({\bf x},t)
\end{equation}
Clearly, this phase factor is unimportant in determining probabilities $|\psi|^2$. Its physical importance becomes clear, when assuming that the electron moves on some orbit around a magnetic field  of magnetic flux $\Phi$ (cf. Fig. \ref{fig-1}). Outside the magnetic field the flux does not exist in the classical view. Quantum mechanically, however, the electron moving outside the magnetic flux tube does indeed feel the presence of the flux phase. 

The decay  of the phase with distance from the flux tube can be derived from the scalar potential $\nabla\phi={\bf A}$ and is given by the surface integral of the magnetic field ${\bf B}$ over the field-line cross section respectively the line integral of the vector potential ${\bf A}$ along the electron orbit
\begin{equation}
\Phi=\int_{\rm cross-sect} {\bf B}\cdot{\rm d}{\bf F}=\int_{\rm orbit} {\bf A}\cdot{\rm d}{\bf s}
\end{equation}
Since the orbit in a scalar potential field can be arbitrarily deformed, the value of the line integral depends only on rotation angle $\theta$ (see Fig.\ \ref{fig-1}), yielding for the potential $2\pi\phi(\theta)=\theta\Phi$. This value, after inserting into the phase factor, gives just
\begin{equation}
\exp\left(-i\theta\frac{e\Phi}{2\pi\hbar}\right)
\end{equation}
which is the result derived by \citet{aharonov1959}. In $\ell$ gyrations of the electron around the field line, the phase factor would increase by $2\pi\ell$, and the wave function $\psi$ would, artificially, become discontinuous. Since this is unphysical, the flux $\Phi=\ell\Phi_0$ in the encircled flux tube \emph{is} a multiple of an elementary \emph{flux quantum}\footnote{This could have been inferred already from simple dimensional reasoning observing that $e\Phi$ has dimension of an action with action quantum $h$.} \citep{aharonov1959}
\begin{equation}
\Phi_0\equiv 2\pi\hbar/e\approx 4.1\times 10^{-15} \quad {\rm Vs}
\end{equation}
This expression \emph{precisely defines a magnetic field line} in magnetically active matter, where the electron is `magnetised', i.e. responds to the presence of the magnetic flux. Of course, magnetic fluxes may form bundles of such elementary flux quanta, but the \emph{smallest possible} magnetic flux in magnetically active matter is given by the ratio $\Phi_0\equiv h/e$ which we propose is the \emph{flux carried by a magnetic field line}. 

\subsection{Field line radius}
This result of \citet{aharonov1959} yields to propose an exact definition of the magnetic field line. The magnetic flux element (quantum) corresponds to a magnetic field of magnitude $B=\Phi_0/\pi\lambda_m^2$which defines a (smallest) magnetic flux tube radius \citep[][unaware of the flux quantum, arrived from different considerations at a similar `magnetic length', not identifying it with field lines]{landau1930}, which is the `radius of a magnetic field line' 
\begin{equation}\label{eq:length}
\lambda_m=\left(\frac{\Phi_0}{\pi B}\right)^\frac{1}{2}=\left(\frac{2\hbar}{eB}\right)^\frac{1}{2}
\end{equation}
It is inversely proportional to the square root of the magnetic field $\sqrt{B}$. Strong magnetic fields correspond to narrow field lines, weak magnetic fields have broader field lines. The field line of a $B=1$ nT field has radius of order $\lambda_m\sim 10^{-3}$ m, which is in the detectable meso-scale domain!%\footnote{Conversely Eq.\ (\ref{eq:length}) can be used to define a largest magnetic field strengths via $B_{\rm P}\lesssim 2\hbar/e\lambda_{\rm P}^2\sim 5\times 10^{54}$ T ,  where $\lambda_{\rm P}$ is the Planck length.}

\begin{figure}[t!]
\centerline{{\includegraphics[width=0.3\textwidth,clip=]{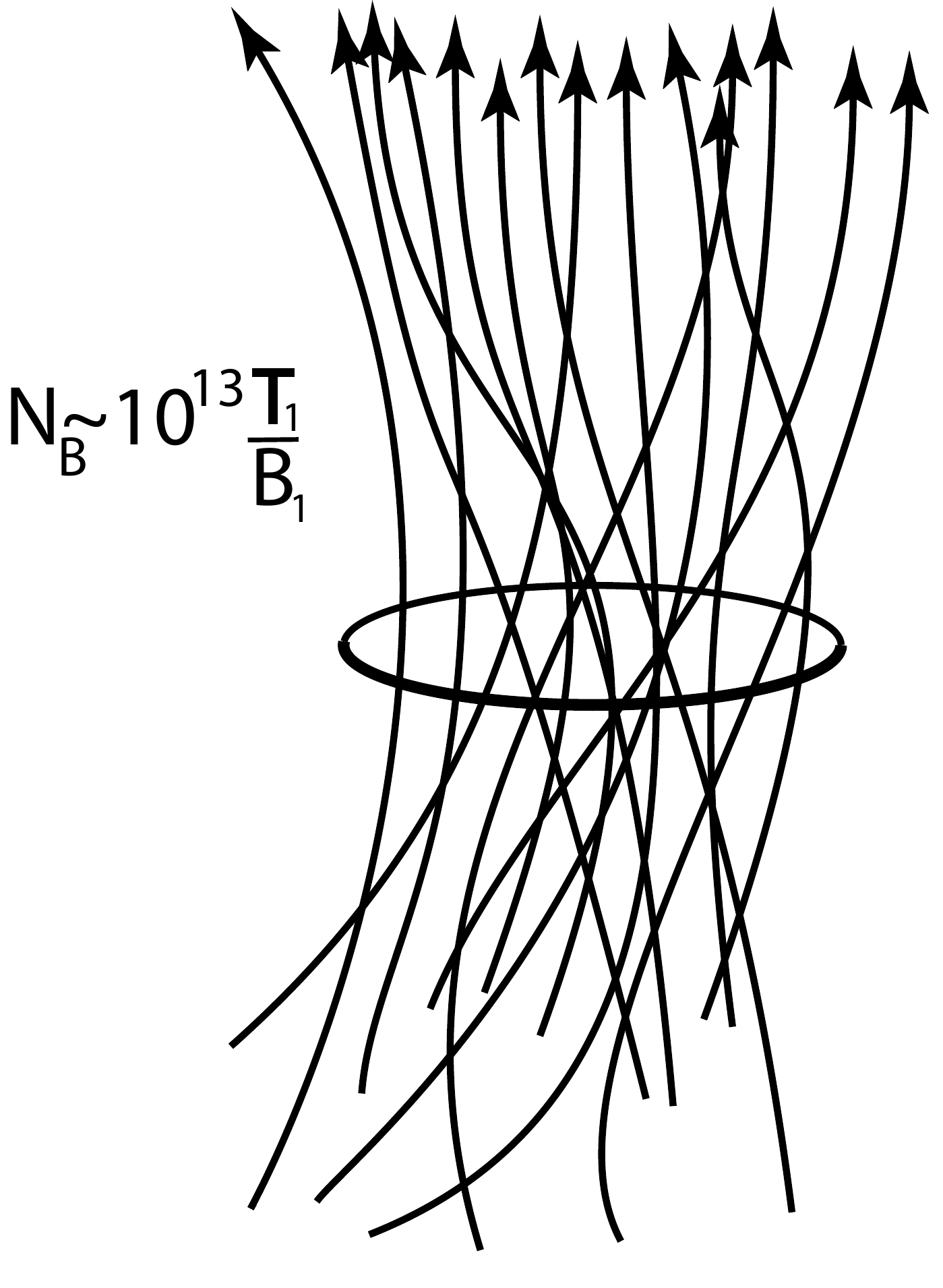}
%\centerline{{\includegraphics[width=0.4\textwidth,clip=]{weibel-fig2a.eps}
}}
\caption[ ]
{\footnotesize A few of the $N_B=T_e/\hbar\omega_{ce}$ magnetic field lines belonging to the bundle of field lines (flux elements of diameter $2\lambda_m$) that are contained in the cross section of an electron gyration ($T_1=1$ eV, $B_1=1$ nT). Because flux can only annihilate in quanta, implying that only \emph{strictly antiparallel} field line segments reconnect, none of these field lines can reconnect, however, even though they may get in touch at some inclination angle.}\label{fig-2}
\vspace{-0.3cm}
\end{figure}

One may calculate the number $N_B$ of magnetic field lines in an electron cyclotron orbit from comparing the area of the gyration circle $\pi r_{ce}^2$ of the electron with the cross section $\pi\lambda_m^2$ of a field line. For an electron of temperature $T_e=mv_e^2/2$ in a magnetic field $B$ this yields a number 
\begin{equation}
N_B= T_e/\hbar\omega_{ce}
\end{equation}
which is the ratio of thermal electron energy to the Landau energy $\hbar\omega_{ce}$ of an electron of cyclotron frequency $\omega_{ce}=eB/m_e$.\footnote{It may be of interest to note that, relativistically, this number becomes $N_B=(m_ec^2/\hbar\omega_{ce})\gamma^2$, i.e. it increases as the square of the relativistic factor $\gamma$.} This number increases with $T_e$ and decreases with magnetic field $B$. In a field of $B=B_1=1$ nT and for an electron temperature of just $T_e=T_{e1}= 1$ eV, this number becomes roughly of the order $N_B\sim 10^{13}$. One electron gyration circle thus contains a huge number of magnetic field lines (see Fig.\ \ref{fig-2}). 

\subsection{`Annihilating' field lines}
Asking what, in principle, will be going on when two oppositely directed field lines, i.e. flux elements containing oppositely directed magnetic fluxes, encounter each other, one realises that the outcome of the encounter depends sensitively on the inclination angle under which the two flux elements, respectively field lines, contact. 

\begin{figure}[t!]
\centerline{{\includegraphics[width=0.375\textwidth,clip=]{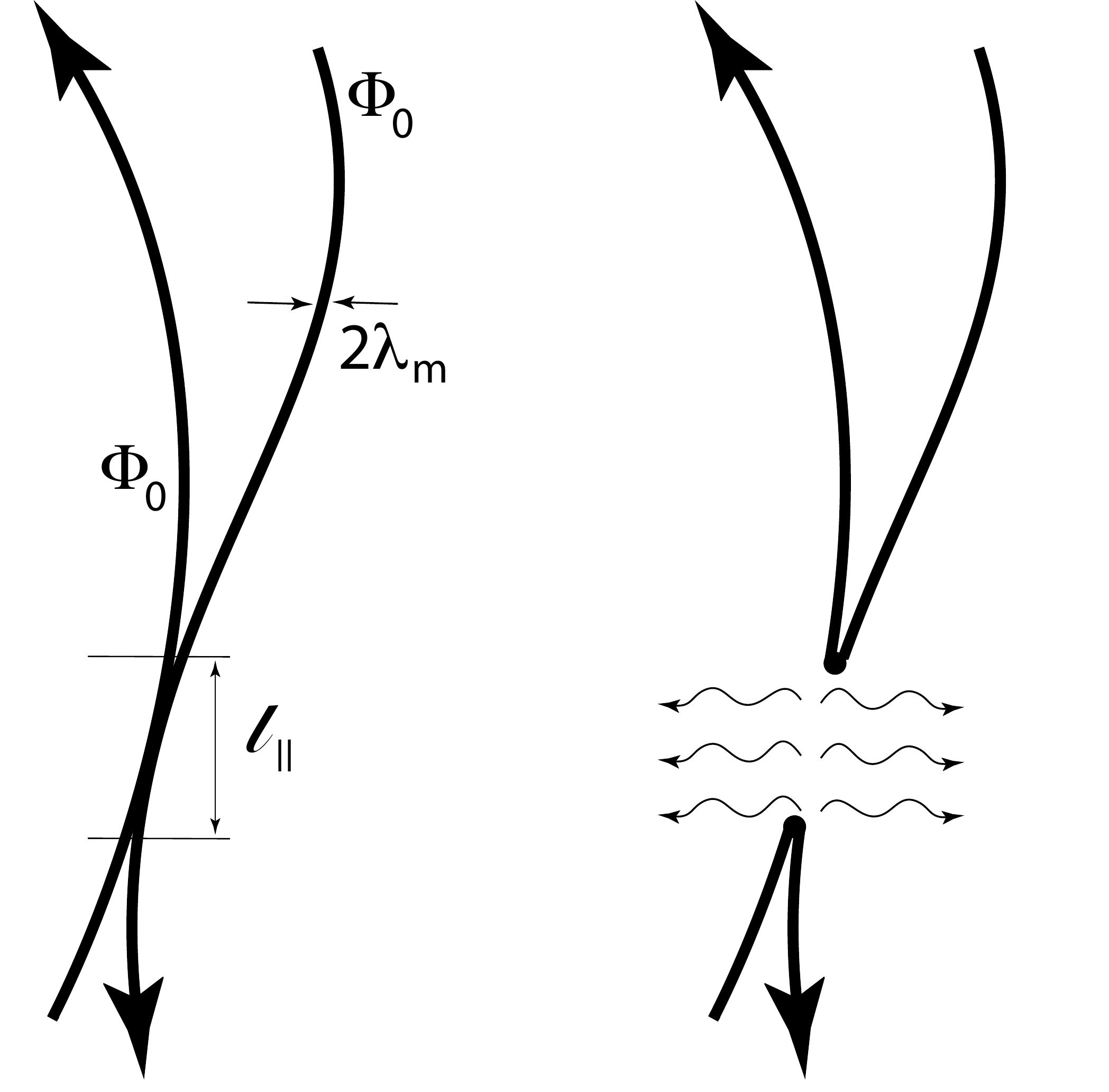}
%\centerline{{\includegraphics[width=0.4\textwidth,clip=]{weibel-fig2a.eps}
}}
\caption[ ]
{\footnotesize Annihilation of the two \emph{strictly antiparallel sections} of two contacting magnetic field lines of diameter $2\lambda_m$, each carrying just one flux quantum $\Phi_0$. Annihilation proceeds over the parallel length $\ell_\|$ only, thereby creating two new field lines each one, as before, carrying just one flux quantum $\Phi_0$. The annihilation releases the magnetic energy stored in the common volume $V_0=2\pi\lambda_m^2\ell_\|$ of the two contacting strictly antiparallel flux tube sections. The released energy heats the plasma locally. The strong bending of the remaining (reconnected) parts of the field lines causes relaxation and stretching of the new field lines but \emph{does not exert any}  forces on the plasma on the narrow microscopic scale of one single field line.}\label{fig-3}
\vspace{-0.3cm}
\end{figure}

Magnetic flux can only be exchanged in quanta $\Phi_0=h/e$. Because of this obvious and undeniable property, annihilation takes only place when the field lines are \emph{precisely} anti-parallel a  certain distance $\ell_\|$ along the elementary flux tubes. This is shown schematically in Fig.\ \ref{fig-3}. One should note that because of this reason any obliquely touching field lines in Fig.\ \ref{fig-2} \emph{cannot} annihilate!

The implication is that, in elementary reconnection events, the \emph{exact} amount of $2\Phi_0$ of magnetic flux will be annihilated. This annihilation happens in a certain time $\Delta t$. Thus the elementary flux annihilation corresponds to the generation of an induced electrical potential difference 
\begin{equation}\label{eq:potential}
|U|={\rm d \Phi}/{\rm d} t\approx 2\Phi_0/\Delta t 
\end{equation}
Multiplying with the elementary charge, the equivalent current density $j= 2B/\mu_0\lambda_m$ surrounding the flux element, and the exactly antiparallel volume $V_0\approx 2\pi\lambda_m^2\ell_\|$, we obtain the total power that is released in such a reconnection event
\begin{equation}\label{eq:power}
P_0\sim \frac{2Bh}{\mu_0e}\ell_\| \sim 10^{-17} (B_1\ell_\|)\quad\mathrm{W}
\end{equation}
It is assumed this total power is converted into kinetic energy of the plasma. Two annihilating field lines of anti-parallel lengths $\ell\sim 1$ m and magnetic field $B=B_1\sim$ 1 nT thus release a power that is just of the order of $\sim10^{-17}$ W. Assuming that the field lines are antiparallel over a length $\lambda_i$ this number becomes the order of $P_0(\lambda_i)\sim 2.5\times 10^{-12} (B_1\sqrt{N_1})$ W. If all the magnetic field lines in an electron gyro-radius would reconnect simultaneously over this length we would multiply by $N_B$,  obtaining 
\begin{equation}
P_0(\lambda_i,N_B)\sim 10~\left(T_{e1}\sqrt{N_1}\right) \quad \mathrm{W}
\end{equation}
In the extreme case when all the electrons in the ion inertial volume would annihilate their field lines one has to multiply by the number of electrons $N\lambda_i^3$, and the total released power would amount to
\begin{equation}
P_\mathit{tot}\sim 10^8~\left(T_{e1}N_1^3\right) \quad \mathrm{W}
\end{equation}
During a substorm of duration $\Delta t\sim 10^3$ s this corresponds to a released energy of 
\begin{equation}
{\cal E}_\mathit{substorm}\sim 10^{11} \quad {\rm J}
\end{equation}
some orders of magnitude higher than measured. Clearly, this is an extreme upper limit of what can be obtained in a reconnection event under the (unreasonable) assumption that it is the \emph{total number} of electrons which (inside the `ion diffusion region' $z<\lambda_i$, i.e. in thin current sheets) are responsible for and contribute to reconnection. 

The released power increases with electron number density $N$ and temperature $T_e$. Apparently, it is independent of the magnetic field $B$, but this is not so because the elementary magnetic flux $\Phi_0$ is contained in the fundamental expression Eq. (\ref{eq:power}) for the released power, and the above expression sums up all contributions of annihilating field lines in one electron gyration circle.

It is of interest to ask how much time the annihilation of two anti-parallel field lines takes. Previously the corresponding amount of energy $e|U|$ was stored in the volume $V_0$ of contacting antiparallel field lines. It thus corresponds to the magnetic energy $V_0B^2/2\mu_0$ stored in the two field-line elements such that we find for the annihilation time of two contacting strictly antiparallel flux elements from Eq. (\ref{eq:potential})
\begin{equation}
\Delta t \sim \frac{h}{e|U|} \approx \frac{e\mu_0}{B\ell_\|}
\end{equation}
This time is extremely short, i.e. flux elements annihilate instantaneously \emph{if and only if} they come into antiparallel contact. Microscopically, from the point of view of field lines (or contacting elementary flux elements), the question of reconnection thus reduces to two problems: ({\it a}) how many flux quanta in an approaching plasma volume element can be turned  antiparallel over a certain length $\ell_\|$, and ({\it b}) how can they be brought into close contact. 

These questions cannot be answered on the base of considering the fluid motion of a magnetised plasma volume. Their answer requires a deeper understanding of the dynamics of the particles in thin (or thinning) current sheets separating globally antiparallel magnetic fields. 

However, the realisation that quantum effects come into play when considering field lines, together with the idea that field lines may reconnect, suggests that the entire problem of reconnection is not an exclusively classical problem but involves the microphysical state of the particles in the electromagnetic field and the quantum nature of their interaction.

\subsection{Magnetic diffusivity}
Before proceeding we briefly note that the existence of a flux quantum implies as well a smallest quantum $D^m_0$ of the magnetic diffusivity such that the diffusion coefficient in a magnetic field of flux $\Phi=\ell\Phi_0$ can be written as
\begin{equation}
D^m_\ell\sim \ell D^m_0, \qquad \ell\in\textsf{N}
\end{equation}
This can be realised when multiplying $\Phi_0$ by the ratio $e/m_e$ of elementary charge to electron mass yielding
\begin{equation}\label{eq-diff}
D^m_0\sim\frac{e}{m_e}\Phi_0=\frac{h}{m_e} \approx 10^{-4} \quad\mathrm{\frac{m^2}{s}}
\end{equation}
which has the correct dimension of a diffusion coefficient: $[D^m]=$ m$^2$s$^{-1}$. Thus, diffusion in a magnetic field proceeds in steps or elementary jumps. 

Correspondingly, the diffusion time $\tau_D^m$ over a certain length $L$ (measured in meters) is then given by 
\begin{equation}
\tau_D^m=10^4(L^2/\ell) \quad\mathrm{s}
\end{equation}

The expression\footnote{Its independence on charge $e$ and magnetic field $B$ identifies it as a general quantum of diffusivity  valid for  stream lines in a fluid.} Eq.\,(\ref{eq-diff}) can be derived in two ways. First assuming with Bohm that the perpendicular displacement during diffusion in one electron gyration time is just of the order of one field line diameter
\begin{equation}
D^m\sim 4\lambda_m^2\omega_{ce}/2\pi
\end{equation}
This yields immediately for
\begin{equation}
D^m\simeq \frac{4\hbar}{\pi m_e}
\end{equation}
Otherwise one may use the definition of the diffusion coefficient through energy
\begin{equation}
D^m\sim 4\pi\epsilon_{\perp}/m_e\omega_{ce}
\end{equation}
where $\epsilon_\perp=\frac{1}{2}\hbar\omega_{ce}$ (see below) is the perpendicular zero point electron energy in the magnetic field $B$. This yields
\begin{equation}
D^m\simeq \frac{2\pi\hbar}{m_e}
\end{equation}
Both expressions agree with $D_0^m$ up to a numerical factor of order $O(1)$.

Since otherwise the magnetic diffusivity is defined through  resistivity $\eta$, one also has
\begin{equation}
D^m_0=\frac{\eta_\perp}{\mu_0}
\end{equation}
an equation which immediately shows that the \emph{perpendicular} ordinary resistivity in a magnetised plasma is itself quantised, $\eta_{\ell\perp}=\ell\eta_{0\perp}$, and cannot be less than its quantum
\begin{equation}
\eta_{\,0\perp}\equiv\frac{\mu_0e}{m_e}\Phi_0=\frac{\mu_0h}{m_e}\approx 10^{-9}\quad{\rm Ohm\ m}
\end{equation}
This fact restricts the ordinary perpendicular conductivity $\sigma_\perp=\sigma_{0\perp}/\ell$ in plasmas to values $\sigma_{0\perp}<10^9$ mho.

\section{Hall physics}
A key observation was the proposal that Hall physics should become invoked in reconnection simply via the demagnetisation of ions on scales below the ion inertial length \citep{sonnerup1979}. Hall physics is essentially two-dimensional even under conditions of three-dimensionality as in the case of a plane thin current sheet. 

The unnecessary (and subsequently misunderstood) complication in this proposal was the inclusion of a field-aligned current closure system which, in principle, does not belong to the Hall currents. The generation of field aligned currents is a peculiarity of the geometry in the geomagnetic tail where the Hall current system is restricted to the region surrounding the magnetic {\sf X}-point during reconnection which along the emanating field connects to the resistive ionosphere thus causing field-aligned currents to flow in order to compensate for the differing electrical conditions. These field-aligned currents are not Hall currents. Rather they are currents which couple the reconnection site to the ionosphere \citep{treumann2009}; they are not an indication of any three-dimensionality of Hall currents even though reconnection as a meso-scale process in the central current layer necessarily is three-dimensional \citep[for a recent discussion cf., e.g.,][]{treumann2010}. Hall currents always form a locally  two-dimensional current system on the meso-scale. 

Noting this two-dimensionality of the Hall current is crucial because it builds the bridge to the above mentioned quantum nature of reconnecting field lines via the quantum Hall effect at very low temperatures in solid state physics. There it was realised that two-dimensional electrons behave quite differently from three-dimensional electrons \citep[cf., e.g.,][for comprehensive reviews]{andrei1997,yoshioka1998}. Two-dimensionality in solid state physics is achieved by artificially (electrostatically) confining the electrons to Hall motion in a plane with the magnetic field being strictly perpendicular to that plane.\footnote{This artificial confinement is necessary in solid state physics. Otherwise the electrons would unavoidably experience scattering at the crystal structure and would not remain two-dimensionally.} In a collisionless plasma in convective motion towards a two-dimensional current sheet two-dimensionality is naturally realised from the very beginning; it needs not to be maintained artificially. Even participation of some electrons in any field aligned current flow does not change this statement because electrons are indistinguishable (identical) particles. 

\subsection{Two-dimensional electrons and Landau levels}
In two-dimensional current sheets, two-dimensionality of the electrons is realised in the Hall region by the orthogonality of the convection electric and magnetic fields. The electric field transports the electrons across the magnetic field by ${\bf E\times B}$-motion constituting, in the ion-inertial region, the Hall current \citep[cf., e.g.,][in addition to the above mentioned observational evidence]{runov2003,nakamura2006}. In the ion-inertial region the ion gyro-radius $r_{ci}$ exceeds the ion-inertial length  $\lambda_i=c/\omega_{pi}$. This is the region where $\beta_i=2\mu_0NT_i/B^2\gtrsim 1$. The ions become non-magnetic here, while the electrons remain to be magnetised carrying the Hall current across the ions. 

Since the plasma is collisionless, it is easy to see from the collisionless electron cross-field drift velocity $|V_{\rm E}|=|E/B|$ and the definition of the Hall current $|J_{\rm H}|=eN|V_{\rm E}|=\sigma_{\rm H}|E|$ that the Hall conductivity is finite and is given by
\begin{equation}
\sigma_{\rm H}= B/eN
\end{equation}
where $N$ is the plasma (or electron) density. The sign of this conductivity is unimportant for the reconnection process and for our discussion here. It suffices to say that the conductivity  is directed perpendicular to the electric convection field and to the ambient magnetic field. Since, however, the electrons behave two-dimensionally, it is not the total electron density which matters but the two-dimensional density $N\ell_\|$, where $\ell_\|$ is the length along the magnetic field within which all electrons behave approximately the same. 

The important point is that such two-dimensional electrons in a magnetic field do not behave classically but behave quantum-mechanically. When solving the Schr\"odinger equation of an electron in a homogeneous magnetic field one finds that their energies are not continuously distributed but occupy evenly spaced Landau energy levels according to the law \citep{landau1930}
\begin{equation}\label{eq:landau}
\epsilon_{\rm L}(p_x, q)=\frac{p_x^2}{2m_e}+\hbar\omega_{ce}(q+{\scriptstyle{\frac{1}{2}}}) \qquad q=1, 2, 3...
\end{equation}
with quantum number $q$. The magnetic field is assumed to point in direction $\pm x$. Hence the perpendicular energy of the electron has a discrete spectrum of equally spaced energy levels with spacing\footnote{Including magnetic mirror geometry causes a split of each Landau level (removing their degeneracy) with quantised energy $\epsilon_b=s\Delta\epsilon_b$, quantum number $s=1,2,...,q$, and energy width $\Delta\epsilon_b=\hbar\omega_b\ll\Delta\epsilon_{\rm L}$, where $\omega_b$ is the electron bounce frequency in the magnetic mirror.} 
\begin{equation}
\Delta\epsilon=\hbar\omega_{ce}
\end{equation}
In weak magnetic fields this spacing is narrow, corresponding to $\Delta\epsilon\sim 10^{-13}$ eV in a magnetic field of $B\sim 1$ nT, increasing proportionally to $B$. For comparison, the Fermi energy in a plasma of electron density $N=10^6$ m$^{-3}$ is $\epsilon_{\rm F}\sim 10^{-17}$ eV, sufficiently small that there is a gap between it and the Landau levels. But in denser plasmas it quickly approaches the lower Landau levels. 

One may note that the field line radius Eq. (\ref{eq:length}) turns out to be the gyro-radius $\lambda_m$ of a lowest Landau level electron of (perpendicular) energy $\Delta\epsilon$. 

\subsection{Filling factors in high temperature conditions}
Landau levels in a thermal plasma are no $\delta$-functions; they have a certain thermal width $\delta\epsilon\sim \frac{1}{2}\epsilon_{\rm L}\sqrt{|b|^2/B^2}$, that depends on the amplitude $|b|$ of the magnetic thermal  fluctuations. Thus, in order for them to not overlap, requires $|b|/B\ll 1$. In the Hall region this condition is satisfied since the non-magnetised ions do not contribute to magnetic fluctuations, and the thermal fluctuations of the magnetic field caused by electrons have indeed very small amplitudes \citep{yoon2007,treumann2010} even though maximising at long wavelengths, $k\to 0$.

Of course, the distribution of the occupation of the Landau levels by the ambient Hall electrons maximises around the (perpendicular) electron temperature $T_{e}$. Nevertheless, there is a number of electrons in the lower Landau levels which have energy less than the spacing between the Landau levels; and these are the electrons which become important in reconnection. Below we will estimate their number density.

Landau level-filling two-dimensional electrons de-magnetise. The Hall conductance becomes independent of the magnetic field $B$ and quantises according to the fundamental Klitzing law 
\begin{equation}\label{eq:hallcond1}
\sigma_{\rm H}= \nu\frac{e^2}{h}
\end{equation}
with the inverse Klitzing constant $e^2/h\equiv\sigma_{\rm K}$, the `quantum of conductivity' \citep[cf., e.g.,][for a still timely account of the quantum Hall effect]{yoshioka1998}. In this case the particle (electron) occupation number $\nu$ of the (lower energy) Landau levels is given by
\begin{equation}\label{eq:hallcond2}
\nu =\frac{\Phi_0}{B}N\ell_\|
\end{equation}
where the right-hand side is just the inverse of the number of field lines in the magnetic field $B$ that are cutting through the surface element. This can be expressed as 
\begin{equation}\label{eq:nu}
\nu\approx 4.14 \frac{N_1\ell_\|}{B_1}
\end{equation}
where $N_1=N/10^6$ and $B_1=B/10^{-9}$ correspond to a plasma of density $N=1$ cm$^{-3}$ and magnetic field $B=1$ nT. Slightly stronger magnetic fields reduce the occupation number while higher densities and longer extensions along the magnetic field increase it. In this respect it is interesting that the theory of the fractional quantum Hall effect suggests large occupation numbers at higher Landau levels with substantially more complicated physics involved.
\begin{figure}[t!]
\centerline{{\includegraphics[width=0.5\textwidth,clip=]{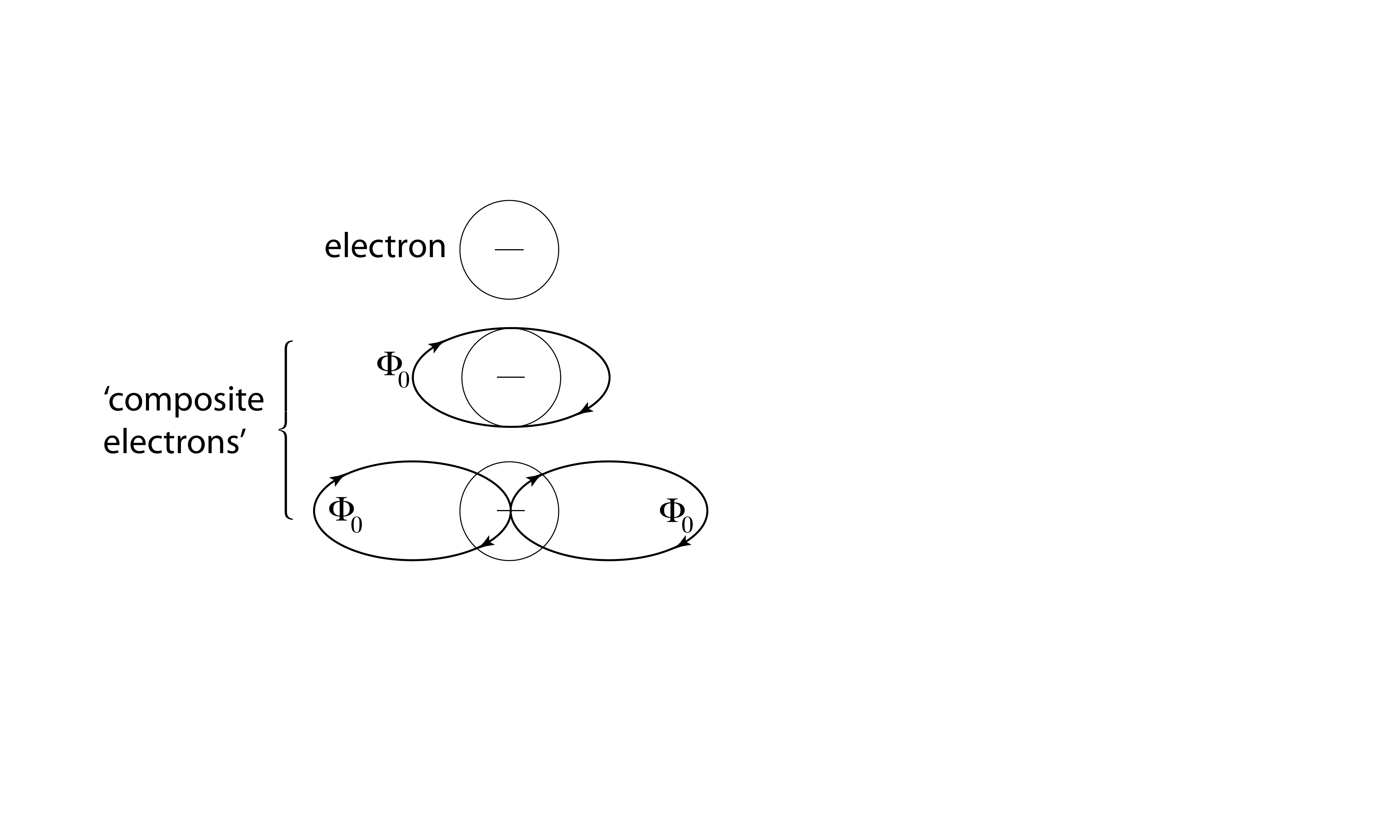}
%\centerline{{\includegraphics[width=0.4\textwidth,clip=]{weibel-fig2a.eps}
}}
\caption[ ]
{\footnotesize A way of attaching flux quanta $\Phi_0$ to an electron thereby transforming the electron into a particle carrying magnetic flux. In the cases shown one and two flux quanta are attached to an electron. The flux quanta occur as closed microscale magnetic vortices which the electron may transport away. One should, however, note that the absorption of a flux quantum by the electron does not imply that a magnetic vortex moves across the plasma. It means that the flux quantum disappears from visibility becoming a quantum property of the electron which is reflected in the different physical and `anyon'-statistical character of the resulting `composite electron'.}\label{fig-4}
\vspace{-0.3cm}
\end{figure}

This number for short extension along the magnetic field is surprisingly small, being comparable to the occupation numbers observed in the quantum Hall effect in solid state physics. On the other hand, for large densities $N_1\gg 1$ and extensions $\ell_\|\gg 1$ m this number becomes very large. For instance, for $\ell_\|\sim \lambda_i\sim 10^3$ km it is $O(10^6)$ or more with most of the electrons being thermal, classical and found only temporarily in one particular Landau level. The thermal spread of their speed lets them stochastically visit many higher Landau levels that are centred around $T_e$ smearing them out. These electrons are of no importance in reconnection. They are `irrelevant' as they remain classical magnetised electrons in the Hall region (the ion-inertial `diffusion' region): they carry classical Hall currents, are magnetised, the Hall currents create quadrupolar Hall magnetic fields in the given {\sf X}-point geometry, thereby twisting the magnetic field slightly. 

All these are secondary effects which have little in common with the micro-scale physics of reconnection. They (`accidentally') happen to occur in a magnetic {\sf X}-point reconnection geometry produced in the proper reconnection process.

\subsection{Level-filling electron number density}
It is only a small number of (collisionless) electrons that escape the classical domain being little or not at all affected by the high temperature. 

These low energy electrons behave the same way as the quantum Hall electrons in solids. One thus must reduce the number density $N_1$ in the above expressions to account only for them and for understanding their effect. These `relevant' electrons may stay for long time in their Landau levels to which they are energetically confined. Confinement will be the case whenever their (perpendicular) thermal energy is of the order or less than the Landau level spacing. 

It was already noted that in an assumed magnetic field of $B= 1$ nT at a temperature of $T_e\sim 1$ eV their transverse thermal energy must be small, less than $\epsilon\lesssim 10^{-13}$ eV. Estimating their number can be done using the Fermi distribution of occupation of energy levels. At the high temperatures of the plasma the Fermi distribution simplifies to the Boltzmann distribution. Since the energy of the electrons under question is $\epsilon\ll T_e$, the Boltzmann exponential function $\exp(-\epsilon/T_e){\rightarrow} 1$ reduces to unity, and the number density in one particular Landau level of quantum number $q$ becomes
\begin{equation}
{\rm d}N_{\rm Lq}(p_x^2)\sim \frac{1}{2}N\sqrt{\frac{\epsilon_q}{T_e}}{\rm d}\left(\frac{\epsilon_q}{T_e}\right)
\end{equation}
where for simplicity we assumed that the parallel energy is $p_x^2/2m_e\sim T_e$, and integration would be only over the Landau energy, such that the parallel energy is constant. Then we obtain for the reduced density of the Landau electrons in Landau level $\epsilon_q=q\hbar\omega_{ce}$, and $q\gg\frac{1}{2}$,
\begin{equation}
\frac{{\rm d}N_{q}}{N}\sim 5\times 10^{-20}\sqrt{q}, \qquad q\ll 10^{13}
\end{equation}
which for $N=N_1$ yields that ${\rm d}N_q\sim 5\times 10^{-14}\sqrt{q}$ m$^{-3}$, substantially less than the ambient thermal `irrelevant' electron density. This is the number density that has to be used in place of $N_1$ in the expression for $\nu$ in Eq. (\ref{eq:nu}) when estimating the filling factor of Landau levels in the quantum Hall regime. 

We can now take a value for $\ell_\|\sim 10^3$ km and find 
\begin{equation}
\nu_q\sim 1.3 \times 10^{-7}\frac{\sqrt{q}}{B_1}\left(\frac{\ell_\|}{10^6~{\rm m}}\right)
\end{equation}
If we chose for the Landau level $q=10^{12}$ we find a canonical filling factor of $\nu_{{\rm log}q=12}\sim 0.13$. 

The lower lying Landau levels also contain those `relevant' electrons, but their numbers are less, and thus their filling factor drops as the square root of $q$. Even though these levels still contribute their effect is thus less. (In fact, one may account for all of them by adding up their contributions. This implies summing over $q$ in the limits $1\leq q\leq q_{\rm max}$. Replacing the sum by an integral then yields an \emph{integrated} contribution of the low lying Landau levels $\propto \frac{2}{3}q^\frac{3}{2}$. Referring to this integrated number, it then suffices to choose the substantially reduced value for $q_{\rm max}=10^8$ in order to obtain  $\int {\rm d}q\nu_q\sim 0.1$, which reproduces the above value for $\nu_q$.) For slightly longer extensions $\ell_\|$ along the field (or larger $q_{\rm max}$) this filling factor comes close to $\nu\sim 1$, in both cases closely corresponding to what is known from quantum Hall effects, in particular from fractal quantum Hall effects \citep[cf., e.g.,][]{nayak2008}. 

The above similarity suggests that under hot dilute (collisionless) plasma conditions in the weak-field reconnection Hall region the processes are like those in the quantum Hall effect at high magnetic field, density and very low temperature. These quantum effects are hidden below the bulk effect of the macro- and mesoscopically dominating classical Hall current that is carried by the `irrelevant' thermal electrons who do not contribute to reconnection, while the `relevant' quantum electrons do, as described below. 

\section{Interpretation: Reconnection scenario}
According to the Pauli exclusion principle for Fermions, Landau levels can be filled with at most two electrons, one spin up, the other spin down. However, the Landau levels Eq. (\ref{eq:landau}) are highly degenerate with degeneracy proportional to the volume $V_0\sim\lambda_i^2\ell_\|$. Since this is a huge number, there is plenty of space in the levels for electrons of both spins, and the level filling will in general be small, with most of the levels being empty, thereby raising the question what the few confined level-filling electrons can do that would be important in reconnection?

One first realises that the electrons that are confined to a Landau level do \emph{not participate} in the thermal motion. Moreover, their contribution to the Hall conductivity is independent of the magnetic field which implies that these electrons \emph{are not magnetised}. This is in contrast to the thermal bulk of the electron distribution which in the ion inertial domain all remain magnetised. 

However, since the confined Landau electrons are also immersed in the magnetic field, their de-magnetisation implies that they \emph{bind some magnetic flux} by absorbing a number $N_{q\Phi}\sim \nu_q^{-1}$ of flux elements as is shown schematically in Fig.\ \ref{fig-4}. The number of absorbed flux elements (or field lines)  is given by the inverse of the filling factor $\nu_q$ of a particular Landau level $q$. 

Clearly, the ability of the confined electrons to absorb magnetic flux changes their physical properties. Such electrons are called \emph{composite electrons} because they consist of electrons that each carry several flux elements $\Phi_0=h/e$. `Composite electrons' belong to the family of \emph{anyons} \citep[for a review cf., e.g.,][]{nayak2008}. They are not subject to Fermi statistics but are described by anyon-statistics, a mixture between bosonic and fermionic statistics that was discovered when interpreting the fractional quantum Hall effect. Their de-magnetisation is just the result of the presence of flux elements (flux quanta $\Phi_0$) that are attached to them. 

That this is so can be realised from consideration of the quantised Hall conductivity Eqs. (\ref{eq:hallcond1},\ref{eq:hallcond2}) where for $\nu=$ const the magnetic field dependence disappears. Its meaning is that the `composite electron' is energetically confined to Landau level $q$.
%However, the mechanism of demagnetisation is more subtle because any \emph{charged} particle necessarily experiences (and reacts to) the presence of a magnetic field unless inertial effects under collisionless conditions (or, otherwise, sufficiently frequent collisions) dominate and inhibit such an interaction. In order to demagnetise the charged particle in the collisionless state (without referring to inertia or collisions) one needs to get rid of its electric charge. 

%For `composite electrons' this is indeed the case: the electrons are somehow `stripped' of their electric charges $e$ in a subtle way. Attaching a flux element to them means multiplying the electrons, i.e. the charge $e$, by the attached number of elementary flux quanta $\nu_q^{-1}h/e$, which just cancels the charge and transforms the electrons into `composite electrons' which do not anymore feel the presence of any ambient magnetic field.
\begin{figure}[t!]
\centerline{{\includegraphics[width=0.45\textwidth,clip=]{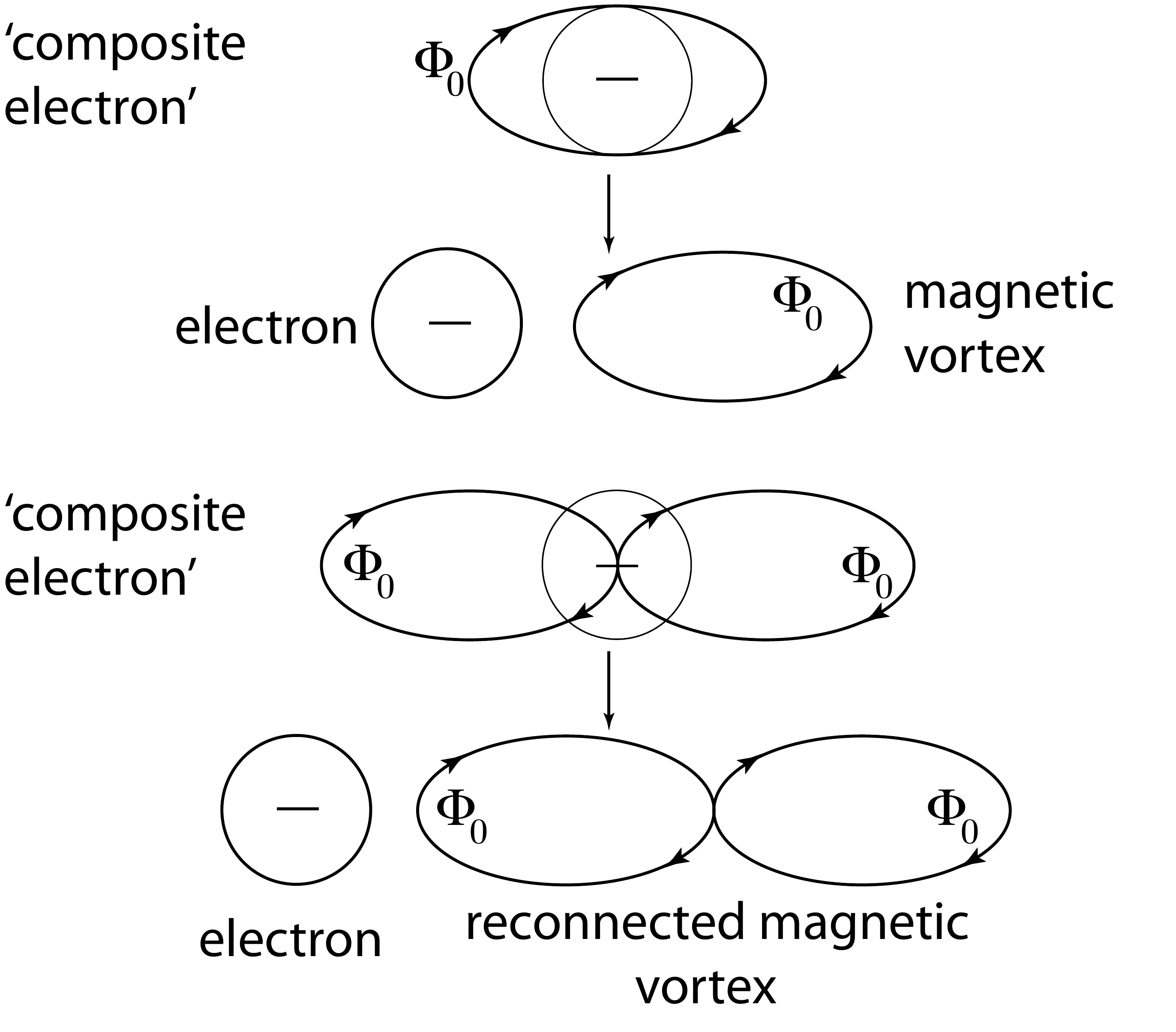}
%\centerline{{\includegraphics[width=0.4\textwidth,clip=]{weibel-fig2a.eps}
}}
\caption[ ]
{\footnotesize Release of attached flux quanta $\Phi_0$ from `composite electrons' in a non-Landau environment (for instance in the non-magnetised centre of the current sheet). The electron regains its normal electron state while creating magnetic vortices in this case. The latter will reconnect, as happens in the second case or when meeting other vortices that have been released from other electrons. This reconnection is a pure vacuum process in which no currents are involved as long as the resulting magnetic structure is of scale smaller than the electron inertial length. After sufficiently many vortices have combined, the inertial scale will be exceeded, and reconnection enters into the domain of classical mesoscale physics. Afterwards the coalescence instability takes over, and reconnection evolves further in the ordinary classical manner.}\label{fig-5}
\vspace{-0.3cm}
\end{figure}

Being non-magnetised, even though still located in the ion inertial domain, these `composite electrons' move freely across the plasma, independently of any electromagnetic interaction. Unlike ordinary thermal Hall electrons which perform a cross-field drift across the orthogonal convection electric and magnetic fields, the  `composite (Landau) electrons'  transport the magnetic flux that is hooked up to them from one place to another. Following their inertia, they cross the ion inertial domain until entering the electron inertial region in the centre of the thin current layer. 

There is, however, one subtlety in this. When the field changes, a `composite electron' leaves its Landau level and might enter another level $q'$. In a continuously though gradually spatially decreasing field $q'>q$. Thus when the `composite electron' of given energy moves toward the electron inertial region it successively steps up the ladder of increasing $q$.  Loaded with magnetic flux quanta, during this motion `composite electrons' transport a certain amount of magnetic flux (and field lines) until they enter the non-magnetic region $|z|<\lambda_e$. Here they do not find any Landau level, become free and are forced to unload their magnetic charge of $\nu^{-1}$ magnetic flux quanta $h/e$ in order to regain their status of ordinary electrons (as schematically shown in Fig.\ \ref{fig-5}). 

Mixing of such `composite electrons' from both sides of the current sheet which carry oppositely directed magnetic flux quanta, causes the necessary mixing of oppositely directed flux quanta (magnetic vortices) for reconnection. It brings the antiparallel flux elements into close contact such that they can annihilate each other and release their stored magnetic energy. Thus, in the present view, the \emph{micro-scale physics of reconnection is a quantum effect} that survives in the domain of meso-scale physics and has observable, even violent, macro-physical consequences. 

It is important to mention a point that might lead to confusion. The transport of flux quanta by the non-magnetised `composite (Landau) electrons' that arise in the Hall domain around a thin current layer implies transport of magnetic field lines into the centre of the current layer. Classically such a transport causes stretching and bending of the transported field line, changing field strength, field geometry, and should thus lead to restoring forces that would ultimately inhibit the motion of the electrons. Apparently this seems to happen as well in the micro-scale domain of the quantum regime when electrons pick up a flux quantum, extract it and transport it away from the location where they have got the load. 

However, though this conclusion is correct in the classical domain, it does not apply to the quantum Hall regime because stretching a flux quantum does not change the quantum. Changing the magnetic field strength implies adding or extracting flux quanta. Loading an electron with a quantum of flux means that the electron \emph{absorbs a microscopic magnetic vortex} as shown schematically in Fig.\ \ref{fig-4}. 

This absorption of a flux quantum by the electron does not imply that a magnetic vortex moves across the plasma. It means that the flux quantum disappears from visibility by becoming a quantum property of the electron which is reflected in the different physical and statistical character of the resulting `composite electron'.

Flux quanta have a fixed value $h/e$ that is completely independent of the geometric form of the corresponding field line. A quantum of magnetic flux (field line) attached to a `composite electron' remains what it is: a quantum of flux of the same value it had when it was picked up by the electron, and it will not change that value when it is transported away and released from the electron in the centre of the current sheet. In other words, attaching a quantum of flux to an electron and transforming the electron into a `composite electron' \emph{removes} the flux element (field line) from the ambient magnetic field. (One may note that this process also somehow removes the electron by transforming it from a charged particle into a `composite electron' which might cause additional unknown effects in the plasma).  Release of the flux element from the `composite electron' \emph{recreates} the flux quantum at the location of its release, i.e. it adds the flux quantum to the environment thereby taking care that no divergence is produced, which means, that it gives birth to a magnetic (field line) vortex. 

Flux quanta can only change when they meet oppositely directed flux quanta and annihilate. This, however, is necessarily the case when they are released from `composite electrons' in the centre of the current sheet because any newly created vortex will contain a segment that is anti-parallel to some other vortex thus allowing for mutual annihilation. 

The annihilation of two antiparallel magnetic vortices in the electron inertial domain, i.e. the newly born (released) flux quanta,  proceeds  without the involvement  of any microscopic currents. Their scales $\lambda_m<N^{-\frac{1}{3}}$ are less than the average interparticle distance. The microscopic vortices therefore occur as though immersed into vacuum, and their interaction and annihilation as well as their reorganisation are pure vacuum effects with annihilation and reorganisation proceeding instantaneously.  The electrons in this region  are unmagnetised and, as argued previously \citep{treumann2010}, become accelerated along the direction of the sheet current by the electric field. Thus  they contribute, via the Weibel instability, to the generation of the macroscopic magnetic field structure. 

\section{Summary and Conclusions}
The above reasoning suggests that the fundamental process of reconnection is of microscopic nature, indeed being a quantum process that leads to profound macroscopic effects in the domain of classical physics. This process proceeds in several steps which we summarise as follows:
\begin{itemize}
\item[*]{It loads electrons with magnetic flux quanta (field line vortices) that are extracted from the domain of Hall current flow. This process of extraction of flux quanta transforms the Landau electrons into `composite electrons'. These are quasi-particles (quasi, because particle-field combinations) and are insensitive to the presence of the magnetic field, i.e. they are, in the language of plasma physics, unmagnetised.} 

\item[*] Being non-magnetic, the `composite electrons' become to some degree independent of the magnetic field and are capable of moving across the plasma. They follow their inertia (or any other non-electromagnetic force like partial pressures, for instance). During their motion they transport the flux quanta that are attached to them into the electron inertial region in the central current layer. Here, no Landau levels exist. When entering the electron-inertial region, the `composite electrons' release the flux quanta, themselves converting back into ordinary electrons. The released flux quanta, on the other hand, form locally closed (divergence-free) small-scale magnetic vortices which subsequently are made available for reconnection.

\item[*] Flowing in from the two sides of the current layer, the magnetic vortices have different senses (right-hand, left-hand), mix and undergo annihilation over their antiparallel parts when coming into contact, thereby releasing the stored magnetic energy which has been extracted from the Hall region by the Landau electrons.

\item[*] Thus the entire process of reconnection implies a gradual (continuous) reduction of magnetic field in the Hall region in terms of extracted magnetic vortices that are attached to the Landau electrons. The subsequent release of these vortices in the electron inertial region, and its local heating/acceleration go exclusively at the  expense of the Hall region. 

\item[*] Since of the magnetic flux vortices (quanta) only their antiparallel parts annihilate, the remaining parts of the vortices rearrange in a necessarily divergence-free manner into the {\sf X}-configuration of the macroscopic magnetic field reconnection structure. This structure implies the production of macroscopic magnetic stresses in a much slower process than annihilation, which cause the observed bulk acceleration of plasma from the {\sf X}-points.   
\end{itemize}

Both the experimental and the theoretical verification/falsification of the proposed scenario and the theory of the micro-scale process that enables collisionless reconnection in thin current sheets will, however, encounter serious difficulties. 

It is obvious from previous theory \citep{sonnerup1979} and observation \citep[][and others]{fujimoto1997,nagai2001,oieroset2001} that thin reconnecting current sheets necessarily include the (classical) Hall effect. This is one necessary precondition for our microscopic mechanism. However, in contrast to the quantum Hall effect in solid state physics which occurs at very low temperatures, the high plasma temperatures will make it difficult to identify the filling effect  of lower energy Landau levels. This is due particularly to the comparatively small number of particles with the required low energies. 

In order to do so one would have to suppress the contribution of thermal electrons to the  Hall conductivity. The classical Hall conductivity increases linearly with magnetic field $B$. One needs to show that this linear increase is occasionally interrupted. This requires extremely precise simultaneous measurements of both $B$ and $N$ during  passages across a current layer in reconnection, such that the classical Hall conductivity can be eliminated. The residual conductivity as function of $B$ should exhibit steps.  

In the differential representation $\Delta\sigma_{\rm H}(B/N)/\Delta B$, which classically for fixed $N$ should be a constant and for pressure equilibrium $N\sim B^2$ decays like $B^{-2}$, the derivative with respect to $B$ is expected to exhibit excursions to low values (in the ideal case to zero). Such excursions, if statistically significant, could possibly indicate the expected Landau confinement and thus the formation of `composite electrons'. 

It would also confirm the occurrence of the quantum Hall effect under the high-temperature dilute collisionless plasma conditions. This unexpected result is, by itself, one of the most exciting suggestions of our theory, which is therefore highly worth being tested by observation and experiment.

Other indications of the effect of `composite electrons' would be an unexplained decrease in magnetic field strength (excluding the Hall field) in the ion-inertial region preceding a reconnection event for some time (which would be of the order of the time electrons need to cross the ion inertial region, i.e. for Landau electrons of energy $\epsilon\sim 0.1T_e$ a time $\tau\sim \sqrt{0.2T_e/\lambda_i^2}$. This decrease in magnetic field strength should be of the same order of magnitude as the field strength in the vicinity of the reconnection site. Also, observation of an irregularly distributed magnetic vortex structure in the central plasma sheet and unusually high magnetic fluctuations around the reconnection site could indicate transport and release of large numbers of field quanta in the electron inertial domain at the centre of the current layer. So far, spacecraft observations seem not to be capable of detecting these micro-scale effects.

On the other hand, numerical simulations in the classical domain will principally miss the underlying micro-scale (quantum) physics as the classical equations of motion on which they are based do not include the proposed effects.  

%\begin{figure}[t!]
%\centerline{{\includegraphics[width=0.4\textwidth,clip=]{weibel-fig2a.pdf}
%\centerline{{\includegraphics[width=0.4\textwidth,clip=]{weibel-fig2a.eps}
%}}
%\caption[ ]
%{\footnotesize The anisotropic-thermal Weibel instability growth rate $\gamma/\gamma_m$, normalised to maximum growth, as function of the normalised wavenumber $k/k_0$ for three different thermal anisotropies. This ratio increases as $A^{-1}$. The vertical line indicates the position of the maximum growing wave number $k_m/k_0$.}\label{weibel-2}
%\vspace{-0.3cm}
%\end{figure}

\begin{acknowledgements}
This research was part of an occasional Visiting Scientist Programme in 2006/2007 at ISSI, Bern. RT thankfully recognises the assistance of the ISSI librarians, Andrea Fischer and Irmela Schweizer. He highly appreciates  the continuous encouragement and support by Andr\'e Balogh, Director at ISSI, and the discussions with Andr\'e Balogh, Stein Haaland and Silvia Perri, all at ISSI, on the present subject, noting the intense though unsuccessful attempts in the cooperation with Stein Haaland to identify signatures of the quantum Hall effect in the observations of dayside magnetopause crossings by the Equator-S and Cluster spacecraft. Lack of particle data in the case of Equator-S and, in the case of Cluster, the for the present purposes completely insufficient resolutions, in particular the resolution of the plasma data, inhibited any unambiguous conclusion.
\end{acknowledgements}


\begin{thebibliography}{ }

\bibitem[Aharonov and Bohm(1959)]{aharonov1959} {Aharonov, Y. and Bohm, D.}: Significance of Electromagnetic Potentials in the Quantum Theory, {Phys. Rev.} 115, 485-491, 1959.

\bibitem[Andrei(1997)]{andrei1997} Andrei, E. Y. (ed.): Two-dimensional electron systems, Springer Series in Physics and Chemistry of Materials with Low-Dimensional Structures, vol.~19 (Springer Verlag, Berlin-Heidelberg-New York, 1997).


\bibitem[Baumjohann et al.(2010)]{baumjohann2010} {Baumjohann, W., Nakamura, R., and Treumann, R. A.}: Magnetic guide field generation in thin collisionless current sheets, {Ann. Geophys.} 28, 789-793, 2010.

\bibitem[Biskamp(2000)]{biskamp2000} Biskamp, D.: Magnetic 
reconnection in plasmas, Cambridge monographs on plasma physics, vol.~3 (Cambridge University Press, Cambridge, UK 2000).

%\bibitem[Buneman(1958)]{buneman1958} {Buneman, O.}: Instability, turbulence, and conductivity in current-carrying plasma: Phys. Rev. Lett. 1,  8-9, 1958, doi: 10.1103/PhysRevLett.1.8.

%\bibitem[Buneman(1959)]{buneman1959} {Buneman, O.}: Dissipation of currents in ionized media: Phys. Rev. 115, 503-517, 1959, doi: 10.1103/PhysRev.115.503.

\bibitem[Drake et al.(2003)]{drake2003} Drake, J.~F., Swisdak, 
M., Cattell, C., Shay, M.~A., Rogers, B.~N., 
and Zeiler, A.: Formation of electron holes and particle energization during magnetic reconnection,  Science 299, 873-877, doi: 10.1126/science.1080333, 2003.

\bibitem[Dungey(1961)]{dungey1961}
{Dungey, J.}: Interplanetary magnetic field and the auroral zones, {Phys. Rev. Lett.} {6}, {47-48, doi: 10.1103/PhysRevLett.6.47}, {1961}.

%\bibitem[Fried(1959)]{fried1959}{Fried, B. D.}: Mechanism for instability of transverse plasma waves, {Phys. Fluids} {2}, {337, doi: 10.1063/1.1705933}, {1959}.

\bibitem[Fujimoto et al.(1997)]{fujimoto1997}
Fujimoto, M., Nakamura, M. S., Shinohara, I., Nagai, T., Mukai, T., Saito, Y., Yamamoto, T., and Kokubun, S.: Observations of earthward stre\-aming electrons at the trailing boundary of a plasmoid, Geophys. Res. Lett. 24, 2893-2896, doi: 10.1029/97GL02821, 1997.

\bibitem[Galeev and Zelenyi(1975)]{galeev1975}
Galeev, A. A. and Zelenyi, L. M.: Metastable states of diffuse neutral sheet and the substorm explosive phase,
{J. Exp. Teor. Phys. Lett.} (JETP Lett.) {22}, {170-172}, {1975}.

\bibitem[Landau(1930)]{landau1930}
Landau, L. D.: Diamagnetismus der Metalle,
{Z. Physik}  {64}, {629-637, doi. 10.1007/BF01397213}, {1930}.

\bibitem[Nagai et al.(2001)]{nagai2001}   Nagai, T., Shinohara, I., Fujimoto, M., Hoshino, M., Saito, Y., Machida, S., and Mukai, T.:   Geotail observations of the Hall current system: Evidence of magnetic reconnection in the magnetotail,   J. Geophys. Res.   106,  25929-25950, doi: 10.1029/2001JA900038,   2001.

\bibitem[Nakamura et al.(2006)]{nakamura2006} Nakamura, R., Baumjohann, W., Asano, Y., Runov, A., Balogh, A., Owen, C. J., Fazakerley, A. N., Fujimoto, M., Klecker, B., and R\`eme, H.: Dynamics of thin current sheets associated with magnetotail reconnection,  J.  Geophys. Res. 111, A11206, doi:10.1029/2006JA011706, 2006.
\newpage
\bibitem[Nayak et al.(2008)]{nayak2008}   Nayak, C., Simon, S. H., Stern, A., Freedman, M., and Das Sarma, S.:   Non-Abelian anyons and topological quantum computation,   Rev. Mod. Phys.   80,  1083-1159, doi: 10.1103/RevModPhys.80.1083,   2008.

\bibitem[Newman et al.(2001)]{newman2001}  {Newman, D. L., Goldman, M. V.,  and Ergun, R. E.}:  Evidence for correlated double layers, bipolar structures, and very-low-frequency saucer generation in the auroral ionosphere,   {Phys. Plasmas} {9}, {2337-2343, doi: 	10.1063/1.1455004}, {2001}.

\bibitem[{\O}ieroset et al.(2001)]{oieroset2001}  { {\O}ieroset, M., Phan, T. D., Fujimoto, M., Lin, R. P., and Lepping, R. P.}:   In situ detection of collisionless reconnection in the Earth's magnetotail,  {Nature}{ 412},   {414-417, doi: 10.1038/35086520},  {2001}.

\bibitem[Parker(1958)]{parker1958}
{Parker, E. N.}:  Sweet's mechanism for merging magnetic fields in conducting fluids, {J. Geophys. Res.} {62}, {509-520, doi: 10.1029/JZ062i004p00509}, {1958}.

{\bibitem[Paschmann et al.(1979)]{paschmann1979}
Paschmann, G., Papamastorakis,I., Sckopke, N., Haerendel, G., Sonnerup, B. U. \"O, Bame, S. J., Asbridge, J. R., Gosling, J. T., Russell, C. T., and Elphic, R. C.: Plasma acceleration at the earth's magnetopause - Evidence for reconnection, {Nature} {282}, {243-246, doi: 10.1038/282243a0}, {1979}.}

{\bibitem[Petschek(1964)]{petschek1964}
Petschek, H. E.: Magnetic field annihilation, in: The Physics of Solar Flares (Proc. AAS-NASA Symp, 28-30 Oct. 1963, GSFC, Greenbelt, MD, W. M. Hess (ed.), Washington, D.C., 1964) pp. {425-439}.}

\bibitem[Runov et al.(2003)]{runov2003} Runov, A., Nakamura, R., 
Baumjohann, W., Zhang, T.~L., Volwerk, M., Eichelberger, H.-U., 
and Balogh, A.: Cluster observation of a bifurcated current sheet, Geophys. Res. Lett. 30, 1036,  doi:10.1029/2002GL016136, 2003. 



%\red{\bibitem[Ricci et al.(2004)]{ricci2004} {Ricci, P., Brackbill, J. U., Daughton, W. and Lapenta, G.}: Collisionless magnetic reconnection in the presence of a guide field, {Phys. Plasmas} {11}, {4102-4114, doi: 10.1063/1.1768552}, {2004}.}

%\bibitem[Sagdeev(1979)]{sagdeev1979}{Sagdeev, R. Z.}: The 1976 Oppenheimer lectures: Critical problems in plasma astrophysics. I. Turbulence and nonlinear waves, II. Singular layers and reconnection, {Rev. Mod. Phys.} {51}, {1-20, doi: 10.1103/RevModPhys.51.11}, {1979}.

%\red{\bibitem[Scholer et al.(2003)]{scholer2003} {Scholer, M., Sidorenko, I., Jaroschek, C. H., Treumann, R. A., and Zeiler, A.}: Onset of collisionless magnetic reconnection in thin current sheets: Three-dimensional particle simulations, {Phys. Plasmas} {10}, {3521-3527, doi: 10.1063/1.1597494}, {2003}.}

\bibitem[Scudder et al.(2008)]{scudder2008}
{Scudder, J. D., Holdaway, R. D., Glassberg, R., and Rodriguez, S. L.}:  Electron diffusion region and thermal demagnetization, {J. Geophys. Res.} {113}, {A10208, doi: 10.1029/2008JA013361}, {2008}.

%\bibitem[Sitenko(1967)]{sitenko1967} Sitenko, A. G.: Electromagnetic Fluctuations in Plasma, Academic Press, New York, 1967.

\bibitem[Sonnerup(1979)]{sonnerup1979} Sonnerup, B. U. \"O.: Magnetic Field Reconnection, in:
Solar System Plasma Physics, Vol III, pp. 45-108, eds. L. T. Lanzerotti, C. F. Kennel and  E. N. Parker, North-Holland, New York, 1979.

\bibitem[Sweet(1957)]{sweet1957}
{Sweet, S.}: The neutral point theory of solar flares, {Proceed. IAU Symp.} {6}, {123-134}, {1958}.

\bibitem[Treumann(2001)]{treumann2001} Treumann, R.~A.: Origin of resistivity in reconnection, Earth, Planets, and Space 53, 453-462, 2001.

\bibitem[Treumann and Baumjohann(1997)]{treumann1997}
Treumann, R. A. and Baumjohann, W.: Advanced Space Plasma Physics, Imperial College Press, London, 1997.

\bibitem[Treumann et al.(2009)]{treumann2009}
{Treumann, R. A., Jaroschek, C. H., and Pottelette, R.}: Auroral evidence for multiple reconnection in the magnetospheric tail plasma sheet, {Europhys. Lett. } 85, {49001, doi: 10.1209/0295 5075/85/49001}, {2009}.

\bibitem[Treumann et al.(2010)]{treumann2010}
{Treumann, R. A., Nakamura, R., and Baumjohann, W.}: Collisionless reconnection: mechanism of self-ignition in thin plane homogeneous current sheets, {Ann. Geophys.} 28, {1935-1943, doi: 10.5194/angeo-28-1935-2010}, {2010}.

\bibitem[Vasyliunas(1975)]{vasyliunas1975} Vasyliunas, V. M.: Theoretical models of magnetic field line merging. I, Rev. Geophys. Space Phys. 13, 303-336, 1975.

%\bibitem[Weibel(1959)]{weibel1959} Weibel, E. S.:  Spontaneously growing transverse waves in a plasma due to an anisotropic velocity distribution, Phys. Rev. Lett. 2, 83-84, doi: 10.1103/PhysRevLett.2.83, 1959.

\bibitem[Yoon(2007)]{yoon2007}   Yoon, P. H.:    Spontaneous thermal magnetic field fluctuations,
Phys. Plasmas 14, 064504-064504-4, doi: 10.1063/1.2741388,   2007.

\bibitem[Yoshioka(2002)]{yoshioka1998}
Yoshioka, D.: The Quantum Hall Effect, Springer Series in Solid-State Sciences, Springer-Verlag, Berlin-Heidelberg-New York, 2002.



%\bibitem[Yoon and Davidson(1987)]{yoon1987}  Yoon, P. H. and Davidson, R. C.:   Exact analytical model of the classical Weibel instability in a relativistic anisotropic plasma, Phys. Rev. A 35, 2718-2721, doi: 10.1103/PhysRevA.35.2718, 1987.

\bibitem[Zeiler et al.(2000)]{zeiler2000} {Zeiler, A., Drake, J. F., and Rogers, B. N.}: Magnetic reconnection in toroidal $\eta_i$ mode turbulence, {Phys. Rev. Lett.} {84}, {99-102, doi: 	 10.1103/PhysRevLett.84.99}, {2000}.

%\red{\bibitem[Zeiler et al.(2002)]{zeiler2002} {Zeiler, A., Biskamp, D., Drake, J. F., Rogers, B. N., Shay, M. A., and Scholer, M.}: Three-dimensional particle simulations of collisionless magnetic reconnection, {J. Geophys. Res. A} {107}, {1230, doi: 10.1029/2001JA000287}, {2002}.}



\end{thebibliography}
\end{document}